\documentclass[aip,jcp,amsmath,amssymb,floatfix,reprint,twocolumn]{revtex4-2}
\usepackage{color,soul}
\usepackage{graphicx}
\usepackage{amsmath, amsthm, amssymb,mathtools}
\usepackage{amsthm}
\usepackage{multirow}
\usepackage[normalem]{ulem}
\usepackage{soul}
\usepackage{hyperref}
\newcommand{\be}{\begin{equation}}
\newcommand{\ee}{\end{equation}}
\newcommand{\bea}{\begin{eqnarray}}
\newcommand{\eea}{\end{eqnarray}}
\newcommand{\eref}[1]{Eq.~(\ref{#1})}%
\newcommand{\fref}[1]{Fig.~\ref{#1}} %
\makeatletter

\newcommand{\Rmnum}[1]{\expandafter\@slowromancap\romannumeral #1@}
\makeatother\usepackage{array, makecell} 

\begin{document}

\title{Mpemba effect in the relaxation of an active Brownian particle in a trap without metastable states}

\author{Apurba Biswas}
\email{apurba.biswas@u-bordeaux.fr}
\affiliation{Laboratoire Ondes et Matière d’Aquitaine, CNRS, UMR 5798, Université de Bordeaux, F-33400 Talence, France}
\author{R. Rajesh} 
\email{rrajesh@imsc.res.in}
\affiliation{The Institute of Mathematical Sciences, C.I.T. Campus, Taramani, Chennai 600113, India}
\affiliation{Homi Bhabha National Institute, Training School Complex, Anushakti Nagar, Mumbai 400094, India}


\begin{abstract}
We explore the role of activity in the occurrence of the Mpemba effect within a system of an active colloid diffusing in a potential landscape devoid of metastable minimum. The Mpemba effect is characterized by a phenomenon where a hotter system reaches equilibrium quicker than a colder one when both are rapidly cooled to the same low temperature. While a minimal asymmetry in the potential landscape is crucial for observing this effect in passive colloidal systems, the introduction of activity can either amplify or reduce the threshold of this minimal asymmetry, resulting in the activity-induced and suppressed Mpemba effect. We attribute these variations in the Mpemba effect to the effective translational shift in the phase space, which occurs as activity is changed.
\end{abstract}

\maketitle

\section{Introduction} 
Mpemba effect refers to the anomalous relaxation phenomena where a system that is initially hotter equilibrates faster than an identical system that is initially cooler, when both systems are quenched to the same low temperature~\cite{Mpemba_1969}. The phenomenon was originally seen experimentally in the freezing of  water when the quench was done across the freezing temperature~\cite{Mpemba_1969,Mirabedin-evporation-2017, vynnycky-convection:2015, katz2009hot, david-super-cooling-1995, zhang-hydrbond1-2014,tao-hydrogen-2017,Molecular_Dynamics_jin2015mechanisms, gijon2019paths}. However, the effect is now known to be a much more general and applicable to the relaxation of any stochastic process, and does not necessarily require a quench across a phase transition. This generality has stimulated the study of the Mpemba effect in a variety of systems both theoretically and experimentally.

Physical systems other than water where the Mpemba effect has now been experimentally observed include magnetic alloys~\cite{chaddah2010overtaking}, polylactides~\cite{Polylactide}, clathrate hydrates~\cite{paper:hydrates},  colloidal systems~\cite{kumar2020exponentially,kumar2021anomalous,bechhoefer2021fresh}, single trapped ion qubit~\cite{shapira2024mpemba}, etc. Numerous theoretical studies on model systems have demonstrated the presence of the Mpemba effect in spin systems~\cite{PhysRevLett.124.060602,Klich-2019,klich2018solution,das2021should,PhysRevE.104.044114,teza2021relaxation},  Markovian systems with few states~\cite{Lu-raz:2017,PhysRevResearch.3.043160}, particles diffusing in a potential~\cite{Walker_2021,Busiello_2021,lapolla2020faster,walker2022mpemba,degunther2022anomalous,biswas2023mpemba,PhysRevE.108.024131,malhotra2024double}, active systems~\cite{schwarzendahl2021anomalous}, spin glasses~\cite{SpinGlassMpemba}, molecular gases in contact with a thermal reservoir~\cite{moleculargas,gonzalez2020mpemba,gonzalez2020anomalous,PhysRevE.104.064127}, quantum systems~\cite{PhysRevLett.127.060401,chatterjee2023quantum,nava2019lindblad,rylands2024microscopic,joshi2024observing,moroder2024thermodynamics,liu2024symmetry,nava2024mpemba,yamashika2024entanglement,chatterjee2023multiple,strachan2024non,ohga2024microscopic,van2024thermomajorization}, systems with phase transitions~\cite{holtzman2022landau,das2021should,zhang2022theoretical,teza2022eigenvalue,chatterjee2024mpemba}, and granular systems~\cite{Lasanta-mpemba-1-2017,Torrente-rough-2019,mompo2020memory,PhysRevE.102.012906,biswas2021mpemba,biswas2022mpemba,megias2022mpemba,biswas2023measure}.

One approach to understanding the Mpemba effect is to study it within a minimal model that exhibits the effect, yet is analytically tractable. In a recent experiment~\cite{kumar2020exponentially}, the Mpemba effect was demonstrated unambiguously in the relaxation of a Brownian particle trapped in an asymmetric double well potential showing that complex inter-particle interactions are not a necessity for the effect. Motivated by this experiment, we studied the overdamped dynamics of a Brownian particle in an asymmetric piecewise linear double-well~\cite{biswas2023mpemba} as well as a single-well potential~\cite{PhysRevE.108.024131}. The linearity of the potential makes it analytically tractable. The existence of the Mpemba effect in the single-well potential with minimal asymmetry  demonstrated that the presence of metastable states is not a necessary condition for the effect~\cite{PhysRevE.108.024131}, in contrast to earlier notion that trapping of the initially colder state in the metastable minima leads to a faster relaxation of the hotter system as suggested in the Refs.~\cite{Lu-raz:2017,PhysRevLett.124.060602,Klich-2019,Walker_2021,schwarzendahl2021anomalous,biswas2023mpemba}. This makes the single-well potential a simple minimal model, devoid of inter-particle interactions as well as multiple minima,  for studying the Mpemba effect.


In this paper, we now ask how the relaxation dynamics of the Brownian particle trapped in the single-well potential is modified in the presence of activity.  Active Brownian particles or active colloids are self-propelling particles that convert chemical energy to mechanical energy thus constantly pumping energy into the system. The presence of activity makes these systems far from equilibrium. The introduction of activity leads to behaviors that are quite distinct from the passive Brownian particle such as the existence of non-Boltzmannian steady states~\cite{RevModPhys.88.045006, malakar2018steady},  wall accumulation~\cite{narinder2019active, volpe2011microswimmers}, activity induced ratchet motion~\cite{reichhardt2017ratchet}, vortices~\cite{bricard2015emergent} and motility induced phase separation~\cite{cates2015motility}. 


There have been a couple of earlier studies of the Mpemba effect in relaxation dynamics in the presence of activity. Along the lines of the experiment of the Mpemba effect in the system of a colloidal particle in an asymmetric double well potential~\cite{kumar2020exponentially}, the role of activity in the same setup was explored numerically in Ref.~\cite{schwarzendahl2021anomalous}. It was shown that activity can induce Mpemba effect in parameter regimes where it is absent for quenches in the passive model and vice-versa for the heating protocol (inverse Mpemba effect). A similar study~\cite{biswas2024mpemba} was done by introducing activity in a discrete three-state Markov process generalizing the relaxation dynamics studied in Ref.~\cite{Lu-raz:2017}. Here, there are energy barriers between the states resulting in activated dominated dynamics. It was shown that activity  leads to unique relaxation phenomena such as the activity induced and suppressed Mpemba effects as well as oscillations in the transients of the relaxation process that are distinct from the passive models.

Both these studies are based on the presence of multiple minima in the underlying energy landscape. Since these minima are not required for observing the Mpemba effect in the passive case, the role of activity is best explored in a single-well potential, thus decoupling  the possible role played by multiple minima. With this motivation, we consider the dynamics of an active Brownian particle diffusing in a single well potential landscape. Although a minimal asymmetry in the potential barriers at its left and right edges leads to the Mpemba effect in the passive model of the Brownian particle, the presence of activity can effectively reduce or enhance the minimal asymmetry. As such it leads to an unique relaxation behavior such as the activity induced and suppressed Mpemba effect when compared to the passive model. The cause of such a phenomena can be mapped to effective translational shift of the existing phase space of the passive system in the presence of activity, leading to the activity induced and suppressed Mpemba effect for a given choice of parameters.

 
The remainder of the paper is organized as follows. We define the model and describe its formalism of the dynamics in Sec.~\ref{model}. Next, we define the Mpemba effect and its necessary criteria in Sec.~\ref{mpemba}. In Sec.~\ref{activity induced and suppressed}, we explore the role of activity in the presence or absence of the Mpemba effect. In Sec.~\ref{phase diagram}, we study the phase diagram of parameter space of the model to understand the underlying significance of activity in the dynamics of the Mpemba effect. Sec.~\ref{conclusion} contains the 
summary of results and a discussion of their implications.




\begin{figure}
\centering
\includegraphics[width= 0.5\columnwidth]{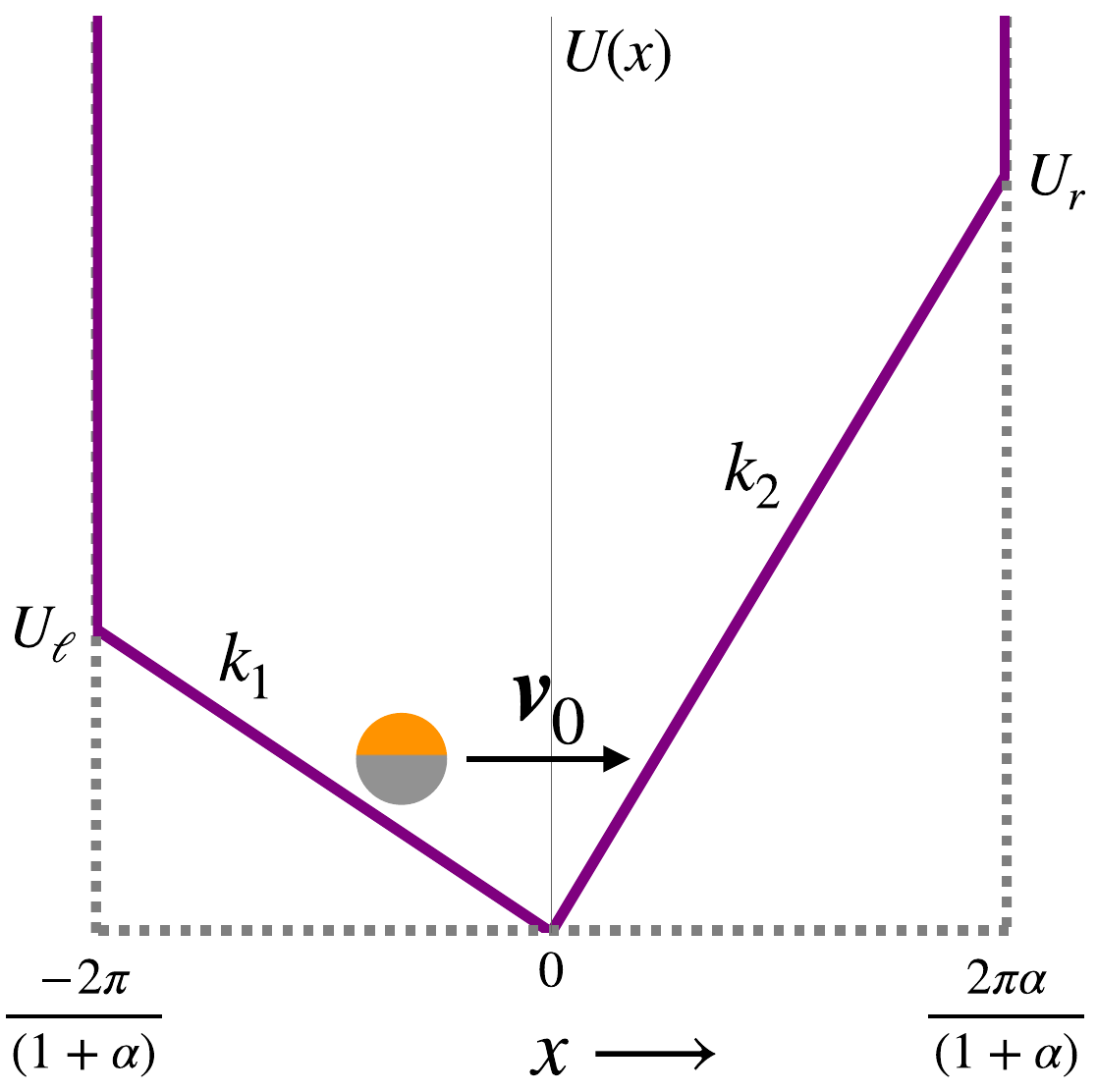} 
\caption{\label{potential shape}Schematic diagram of the system of an active particle in a piecewise linear potential. The activity of the colloidal particle is characterized by its persistence velocity $\boldsymbol{v}_0$. The boundaries of the potential are situated at $\frac{-2\pi}{(1+\alpha)}$ and $\frac{2\pi \alpha }{(1+\alpha)}$ with the total length of the domain fixed at $2 \pi$. The parameters $k_1$ and $k_2$ refer to the various slopes, $\alpha$ denotes the asymmetry factor for the widths of left and right domains, and $U_{\ell}$ and $U_r$ refer to the potential heights at the left and the right boundaries.}
\end{figure}

\section{\label{model}Model and formalism}
We consider an active Brownian particle placed in a one dimensional potential, $\tilde{U}(\tilde{x})$ in the presence of a thermal environment characterized by damping $\tilde{\gamma}$ and noise $\tilde{\eta}$ with the statistics $\langle \tilde{\eta}(t) \rangle=0$ and $\langle \tilde{\eta}(\tilde{t}) \tilde{\eta}(\tilde{t}') \rangle= 2 \gamma k_B \tilde{T}_b \delta (\tilde{t}-\tilde{t}')$. Here, $\tilde{T}_b$ is the temperature of the thermal bath and $k_B$ is the Boltzmann's constant. The equation of motion of the active particle in the overdamped approximation, where the damping $\gamma$ is large compared to the mass of the particle, is given by the Langevin equation~\cite{schwarzendahl2021anomalous}:
\be
 \frac{d\tilde{x}}{d\tilde{t}}=\tilde{v}_0 n-\frac{1}{\gamma}\frac{d\tilde{U}}{d\tilde{x}}+\tilde{\eta}(\tilde{t}), \label{langevin eqn}
\ee
where $\tilde{v}_0$ is the self-propulsion speed and $n$ denotes its direction of motion where $n=\pm1$ for $\pm x$-direction. The value of $n$ flips stochastically with a waiting time $\tilde{t}_p$ drawn from an exponential distribution $p(\tilde{t}_p)=\tilde{\tau}^{-1}_p e^{-\tilde{t}_p/\tilde{\tau}_p}$ with $\tilde{\tau}_p$ being the persistence time. 

We denote the probability density of the particle propelling to the right by $P_r(\tilde{x},\tilde{t})$ and that for the left with $P_l(\tilde{x},\tilde{t})$. The equations of motion in terms of the probability densities are given by~\cite{schwarzendahl2021anomalous}:
\bea
\frac{\partial P_r}{\partial \tilde{t}} &=  
\frac{1}{\gamma}\frac{\partial}{\partial \tilde{x}}\Big[ \frac{d\tilde{U}}{d\tilde{x}} P_r \Big]-\tilde{v}_0\frac{\partial P_r}{\partial \tilde{x}} + \frac{k_B \tilde{T}_b}{\gamma} \frac{\partial^2 P_r}{\partial \tilde{x}^2} -\frac{P_r}{\tilde{\tau}_p}  + \frac{P_l}{\tilde{\tau}_p} ,
&\label{Pl}
\eea 

\bea
\frac{\partial P_l}{\partial \tilde{t}} &=\frac{1}{\gamma}\frac{\partial}{\partial \tilde{x}}\Big[ \frac{d\tilde{U}}{d\tilde{x}} P_l \Big]+\tilde{v}_0\frac{\partial P_l}{\partial x} + \frac{k_B \tilde{T}_b}{\gamma} \frac{\partial^2 P_l}{\partial \tilde{x}^2} -\frac{P_l}{\tilde{\tau}_p}  + \frac{P_r}{\tilde{\tau}_p} . \label{Pr}
\eea 
%
We now introduce the following dimensionless variables: $x=(2\pi/L)\tilde{x}$, $T=\tilde{T}/\tilde{T_b}$, $U=\tilde{U}/(k_B T_b)$, $t=\tilde{t}/\tau_p$, $v_0=(2\pi \tilde{v}_0 \tau_p)/L$ and $\tau_p=(4 \pi^2 k_B T_b \tilde{\tau_p})/(\gamma L^2)$. In terms of the dimensionless variables, the time evolution of the probability densities in Eqs.~(\ref{Pl}) and (\ref{Pr}) can be rewritten as: 
\bea
\frac{\partial P_l}{\partial t} &=  v_0 \frac{\partial P_l}{\partial x} + \tau_p \frac{\partial}{\partial x}\Big[ \frac{dU}{dx} P_l\Big] + \tau_p \frac{\partial^2 P_l}{\partial x^2}- P_l + P_r, \label{Pl dimensionless} \\
\frac{\partial P_r}{\partial t} &=  -v_0 \frac{\partial P_r}{\partial x} + \tau_p \frac{\partial}{\partial x}\Big[ \frac{dU}{dx} P_r\Big] + \tau_p \frac{\partial^2 P_r}{\partial x^2}- P_r + P_l. \label{Pr dimensionless}
\eea 
The densities for the total occupation probability, $P(x,t)$, and for the polarization, $Q(x,t)$, are defined as
\bea
P(x,t)=P_r(x,t)+P_l(x,t),\\
Q(x,t)=P_r(x,t)-P_l(x,t).
\eea
The quantity $P(x,t)$ denotes the probability density of finding the active particle at position $x$ and time $t$ irrespective of its bias to move in positive or negative $x$ direction. On the other hand, the polarization density $Q(x,t)$ denotes the preferential bias at position $x$ and time $t$ to move in the positive $x$-direction over the negative direction. The quantities $P(x,t)$ and $Q(x,t)$ evolve in time as 
\begin{flalign}
&\frac{\partial P}{\partial t}=-v_0 \frac{\partial Q}{\partial x} + \tau_p\frac{\partial}{\partial x}\Big[ \frac{dU}{dx} P\Big]  + \tau_p \frac{\partial^2 P}{\partial x^2}, \label{eq P} \\
&\frac{\partial Q}{\partial t}=-v_0 \frac{\partial P}{\partial x} + \tau_p\frac{\partial}{\partial x}\Big[ \frac{dU}{dx} Q \Big]  + \tau_p \frac{\partial^2 Q}{\partial x^2}-2Q. \label{eq Q}
\end{flalign}

We consider a single well potential that is piece-wise linear. The choice of such a potential gives an analytically tractable model for the system of passive Brownian particle, and motivates us to consider the same in the case of active system in order to make a direct comparison of the key results with the passive system for a given parameter space. The boundaries of the well are situated at $(x_{min},x_{max})$ with $x_{min}=- (2\pi)/(1+\alpha)$ and $x_{max}=(2 \pi \alpha)/(1+\alpha)$ such that the total length of the domains is fixed at $2\pi$ and the parameter $\alpha$ determines the asymmetry between the widths of the left and right domains. The construction of the potential in this way helps in reducing the number of variables in the model. As such, the parameters characterizing the  configuration of the potential are: $U_{\ell}$, $U_r$ and $\alpha$. The shape of the potential is shown in Fig.~\ref{potential shape}, and is described quantitatively as
\be
U(x)=
\begin{cases}
U_{\ell}+k_1 (x-x_{min}), & x_{min} < x < 0, \\
k_2 x , & 0 < x < x_{max},  \label{potential form}
\end{cases}
\ee
where $k_1=U_{\ell}/x_{min}$, $k_2=U_r/ x_{max}$, $\alpha$, $U_{\ell}$ and $U_r$ are constants.

The equations (\ref{eq P}) and (\ref{eq Q}) can be written in a concise matrix form as
\be
\frac{\partial \boldsymbol{P}}{\partial t}=\boldsymbol{\mathcal{L}}\boldsymbol{P}, \label{eq matrix L}
\ee
where the vector $\boldsymbol{P}=(P,Q)^{Tr}$, where $Tr$ represents transpose and matrix
\begin{align}
&\boldsymbol{\mathcal{L}}\!=\nonumber \\
&\!\left[\begin{array}{cc} 
\tau_p \frac{d^2U}{dx^2} +\tau_p\frac{dU}{d x}\frac{\partial}{\partial x}  + \tau_p \frac{\partial^2 }{\partial x^2} & -v_0\frac{\partial}{\partial x}\\
-v_0 \frac{\partial}{\partial x} & \tau_p \frac{d^2U}{dx^2} +\tau_p\frac{dU}{d x}\frac{\partial}{\partial x}  + \tau_p \frac{\partial^2 }{\partial x^2}-2
\end{array}\right].
\label{eq:Lmatrix}
\end{align} 
Since the potential diverges at the boundaries, no flux  boundary condition is implemented and is given by
\be
\boldsymbol{j}(x_{min})=\boldsymbol{j}(x_{max})=0,
\ee
where the probability current/ flux is given by
\begin{equation}
\boldsymbol{j}\!=\!\left(\begin{array}{c} 
-\tau_p \frac{dU}{dx} P - \tau_p \frac{\partial P }{\partial x} +v_0 Q\\
-\tau_p \frac{dU}{dx} Q - \tau_p \frac{\partial Q }{\partial x} +v_0 P
\end{array}\right).
\label{eq:Jmatrix}
\end{equation}

The solution for \eref{eq matrix L} is obtained using eigenspectrum decomposition method as
\be
\boldsymbol{P}(x,t)=\boldsymbol{\pi}(x,T_b) + \sum_{i\geq 2} a_i (T,T_b) \boldsymbol{v}_i(x) e^{\lambda_i t}, \label{full time ev}
\ee
where, 
$\boldsymbol{v}_i$ are the right eigenfunctions of $\boldsymbol{P}(x,t)$, $\lambda_i$ are the eigenvalues which follow the order $\lambda_1=0>\lambda_2>\lambda_3 \ldots$ and $\boldsymbol{\pi}(x,T_b)$ is the final steady state distribution corresponding to $\lambda_1=0$. 
In order to compute the coefficient $a_i$, we consider the case of $t=0$ such that
\bea
\boldsymbol{P}(x,0)=\boldsymbol{\pi}(x,T_b) + \sum_{i\geq 2} a_i (T,T_b) \boldsymbol{v}_i(x).
\eea
Note that $\boldsymbol{P}(x,0) \equiv \boldsymbol{\pi}(x,T)$ and is the initial steady state distribution at a temperature $T \neq T_b$ where the system is initially prepared to before being quenched to the final  bath temperature $T_b$.  Then the coefficients $a_i(T,T_b)$ are given by the inner product of the left eigenfunctions $\boldsymbol{u}_i(x)$ (solved for bath temperature $T_b$) with the initial steady state distribution $\boldsymbol{\pi}(x,T)$ as
\bea
a_i(T,T_b)=\frac{\langle \boldsymbol{u}_i|\boldsymbol{\pi}(x,T)\rangle}{\langle \boldsymbol{u}_i| \boldsymbol{v}_i\rangle}.
\eea

We now discuss about the computation of the left eigenfunctions $\boldsymbol{u}(x)$ and the initial steady state distribution $\boldsymbol{\pi}(x,T)$. To compute the left eigenfunctions $\boldsymbol{u}(x)$, one needs to work with the adjoint operator $\boldsymbol{\mathcal{L}}^\dagger$ of $\boldsymbol{\mathcal{L}}$. In order to compute $\boldsymbol{\mathcal{L}}^\dagger$, we consider two test functions $\boldsymbol{\psi}(x)$ and $\boldsymbol{\phi}(x)$ and do the following exercise:
\bea
\langle \boldsymbol{\psi}(x)| \boldsymbol{\mathcal{L}} \boldsymbol{\phi}(x) \rangle=\int^{x_{max}}_{x_{min}}  dx \boldsymbol{\psi}^*(x) \boldsymbol{\mathcal{L}} \boldsymbol{\phi}(x).
\eea

After doing the integration by-parts for the components of the operator $\boldsymbol{\mathcal{L}}$ and rearranging, we obtain
\begin{flalign}
&\langle \boldsymbol{\psi}(x)| \boldsymbol{\mathcal{L}} \boldsymbol{\phi}(x) \rangle \nonumber\\
&=\langle \boldsymbol{\psi}(x) \boldsymbol{\mathcal{L}}^\dagger|  \boldsymbol{\phi}(x) \rangle - \left[ \boldsymbol{\psi}. \boldsymbol{j} + \frac{k_B T_b}{\gamma} \frac{\partial \boldsymbol{\psi}}{\partial x}. \boldsymbol{\phi} \right]^{x_{max}}_{x_{min}},
\end{flalign}
where the adjoint operator $\boldsymbol{\mathcal{L}}^\dagger$ is given by
\begin{equation}
\boldsymbol{\mathcal{L}}^{\dagger}\!=\!\left[\begin{array}{cc} 
-\tau_p\frac{dU}{d x}\frac{\partial}{\partial x}  +\tau_p \frac{\partial^2 }{\partial x^2} & -v_0\frac{\partial}{\partial x}\\
-v_0 \frac{\partial}{\partial x} & -\tau_p\frac{dU}{d x}\frac{\partial}{\partial x}  + \tau_p \frac{\partial^2 }{\partial x^2}-2
\end{array}\right],
\label{eq:L adjoint matrix}
\end{equation} 
and it clearly obeys the Neumann boundary conditions given in terms of its eigenfunctions $\boldsymbol{\psi}(x)$ as 
 \bea
 \frac{\partial \boldsymbol{\psi}}{\partial x}=0~\text{at} ~x=~x_{min},~x_{max}.
 \eea
 

The initial steady state distribution $\boldsymbol{\pi}(x,T)$ at temperature $T$ is obtained by numerically solving the coupled differential equations \eref{eq P} and \eref{eq Q} with the time derivative set to zero and with their appropriate boundary conditions. Although the coupled differential equations are not analytically solvable even with the piece-wise linear, but the solutions are numerically exact.

 \section{\label{mpemba}The Mpemba effect}
 With the above framework, we aim to study the consequence of the activity on the anomalous relaxation dynamics, namely the Mpemba effect. It refers to the faster relaxation of an initially hotter system compared to an initially warmer system when both are quenched to a common final state characterized by an even colder temperature. For the current system under consideration, the temperature of the heat bath sets the temperature of the system. However, note that due to the presence of external active forces, the system has a non-equilibrium steady state instead of the equilibrium state and it is characterized by the distribution $\boldsymbol{\pi}(x,T)$ corresponding to some choice of the active parameters $v_0$, $\tau_p$ and temperature $T$ of the heat bath. Keeping all the parameters the same, the quench from one steady state to another is done by changing the temperature of the bath.

Upon quench, the time evolution of the probability distribution $\boldsymbol{P}(x,t)$ is given by \eref{full time ev}. Since at large times, only the first non-zero largest eigenvalue dominates, the condition for the Mpemba effect can be obtained from the long time limit of the evolution \eref{full time ev} which is given by
\bea
\boldsymbol{P}(x,t)\simeq \boldsymbol{\pi}(x,T_b) +  a_2 (T,T_b) \boldsymbol{v}_2(x) e^{\lambda_2 t}, \label{large time behavior of P}
\eea
where $a_2$ is given by
\bea
a_2=\frac{\langle \boldsymbol{u}_2|\boldsymbol{\pi}(x,T)\rangle}{\langle \boldsymbol{u_2}| \boldsymbol{v_2}\rangle}. \label{a2}
\eea

Now, let us consider two identical systems which are prepared at two different initial steady states corresponding to the choice of a hot and warm bath temperatures $T_h$ and $T_c$ respectively. As both the systems are quenched to the common steady state of the bath temperature $T_b$ such that $T_h>T_c>T_b$ then the Mpemba effect is said to exist if
\begin{equation}
|a_2(T_h,T_b)|<|a_2(T_c,T_b)|. \label{condition}
\end{equation}

It is so because with the above condition, the distribution $\boldsymbol{P}(T_c, t)$ of the initially cold system lags behind the distribution $\boldsymbol{P}(T_h, t)$ of the initially hot system leading to the Mpemba effect. Thus, the overall task reduces to computing the coefficient $a_2(T,T_b)$ in \eref{a2}.

\begin{figure}
\centering
\includegraphics[width=\columnwidth]{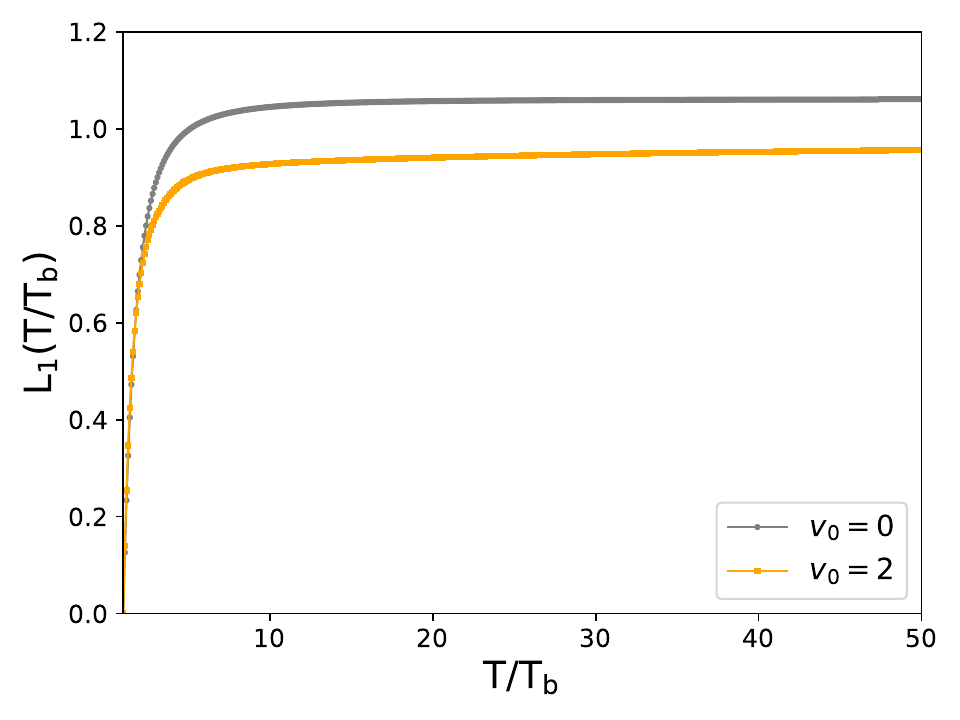} 
\caption{\label{L1 vs T}Variation of the distance function $L_1$ as a function of the initial temperatures $T$. The final quenched temperature corresponds to $T_b=1$. Monotonic rise of $L_1(T)$ with $T$ implies that $L^h_1>L^c_1$ for the initially hot and cold systems with temperatures $T_h$ and $T_c$ respectively and the result is independent of the passive or active cases of the model as shown here for $v_0=0$ and $v_0=2$.}
\end{figure}

Equivalently, the relaxation process can also be described in terms of distance from steady state function, $D[\boldsymbol{P}(t),\boldsymbol{\pi}(T_b)]$ which measures the instantaneous distance of a distribution $\boldsymbol{P}(x,t)$ from the final steady state distribution, $\boldsymbol{\pi}(T_b)$. There exists numerous well defined distance measures in literature such as Kullback-Leibler (KL) divergence~\cite{Lu-raz:2017,kumar2020exponentially,Busiello_2021}, entropic distance, $L_1$-norm and so on. Thus, in terms of the distance function, the initially hot and the cold system prepared at temperatures $T_h$ and $T_c$ respectively are denoted by the inequality $D[\boldsymbol{\pi}(T_h), \boldsymbol{\pi}(T_b)]>D[\boldsymbol{\pi}(T_c),\boldsymbol{\pi}(T_b)]$ for $T_h > T_c$. If it is followed by $D[\boldsymbol{P}_h(t),\boldsymbol{\pi}(T_b)]<D[\boldsymbol{P}_c(t),\boldsymbol{\pi}(T_b)]$ at a later time $t$, we state that  the Mpemba effect exists.

In this paper, we discuss the $L_1$-norm for its ease of convenience to compare with the discussion in terms of $a_2$  measure. It is defined as 
\bea
\begin{split}
D[\boldsymbol{P}(t),\boldsymbol{\pi}(T_b)] &\equiv L_1(t)=\int dx |\boldsymbol{P}(x,t)-\boldsymbol{\pi}(x,T_b)|. \label{L1 measure-def}
\end{split}
\eea

We will now show that the use of measure $|a_2(T,T_b)|$ is equivalent to using any other distance measure, namely $L_1$-measure. For that purpose, we need to understand the behavior of $L_1$ at time $t=0$ for different temperatures and also how it behaves at large times. First, we show that at $t=0$, $L_1(T)$ is a monotonically increasing function of temperature $T$ irrespective of the passive or active model as shown in Fig.~\ref{L1 vs T}. As a result, for two initially hot and cold systems with temperatures $T_h$ and $T_c$ with $T_h>T_c$, it means that $L^h_1>L^c_1$. At large times, using Eq.~(\ref{large time behavior of P}), we can write for $L_1(t)$ as
\bea
L_1(t)\simeq  |a_2 (T,T_b)| \int dx~  | \boldsymbol{v}_2(x) e^{\lambda_2 t} |.
\eea

Thus, if $|a_2(T_h, T_b)|<|a_2(T_c, T_b)|$, then $L^h_1< L^c_1$, i.e., $hot$ system is closer to the final steady state compared to the initial $cold$ system. From the discussion, it is quiet clear that the measure $a_2(T, T_b)$ is sufficient to describe the Mpemba effect for the present model.

\section{\label{activity induced and suppressed}Role of activity in the Mpemba effect}
With the above formalism, we look at the effect of active parameters of the model: $v_0$ and $\tau_p$, in the existence of the Mpemba effect. Note that $v_0=0$ corresponds to the model of passive Brownian particle and $v_0\neq 0$ corresponds to that of the active particle while keeping every other parameters of the potential and bath the same for both the cases. Since the parameters $v_0$ and $\tau_p$ are coupled to each other, for the simplicity of the analysis, we shall set $\tau_p$ as unity unless otherwise mentioned and increase in activity will henceforth refer to increase in $v_0$.  To that end, we explore various configurations of the external potential and ask if the presence of activity ($v_0\neq 0$) helps or suppress the existing phase space of the Mpemba effect that is already known for the passive scenario of the model.  

\subsection{Mpemba effect suppressed due to activity}
\begin{figure}
\centering
\includegraphics[width=\columnwidth]{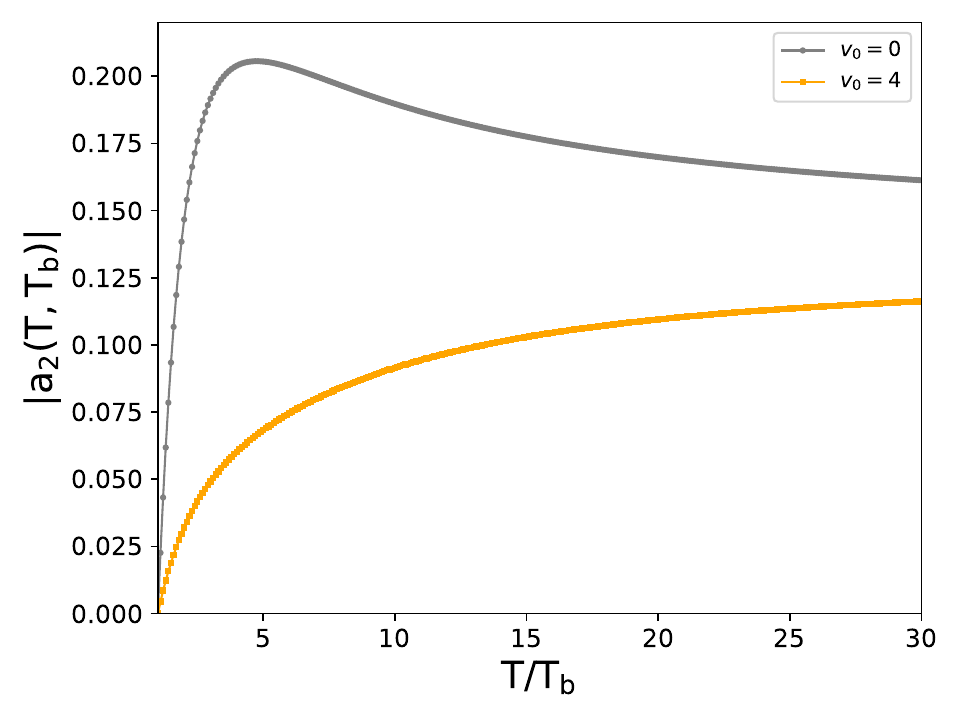} 
\caption{\label{a2 vs T}Increase in activity suppressing the Mpemba effect. Variation of $|a_2(T,T_b)|$ with $T$ for the chosen set of parameters: $U_{\ell}$=4, $U_r$=10, $\gamma=1$, $\tau_p=1$ and for the choices of the persistence velocities: $v_0=0$ and $v_0=4$. The final quenched temperature corresponds to $T_b=1$. With the increase in $v_0$, the Mpemba effect vanishes.}
\end{figure}

We first consider a case where the Mpemba effect already exists in the passive scenario of the model. In our example, it corresponds to making a choice of the potential parameters $U_{\ell}=4.0$, $U_r=10.0$ and $\alpha=1.0$ and the bath temperature corresponding to the final steady state is set at $T_b=1.0$. The presence of the  Mpemba effect in this parameter space is confirmed by the  non-monotonic variation of the coefficient $|a_2(T,T_b)|$ with $T$ for $v_0=0$ as shown in \fref{a2 vs T}. 

Now with the introduction of the activity parameter $v_0$ and keeping the same configuration of the potential, the Mpemba effect persists over a wide range of initial temperatures. However, as we further increase $v_0$, the Mpemba effect vanishes as characterized by the monotonic rise of $|a_2(T,T_b)|$ with $T$ as shown in \fref{a2 vs T} for $v_0=4$. Thus, with the example that we considered, increase in activity or in other words increasing the persistence velocity of the particle suppresses the Mpemba effect.



\subsection{Intermediate activity can induce Mpemba effect}
\begin{figure}
\centering
\includegraphics[width=\columnwidth]{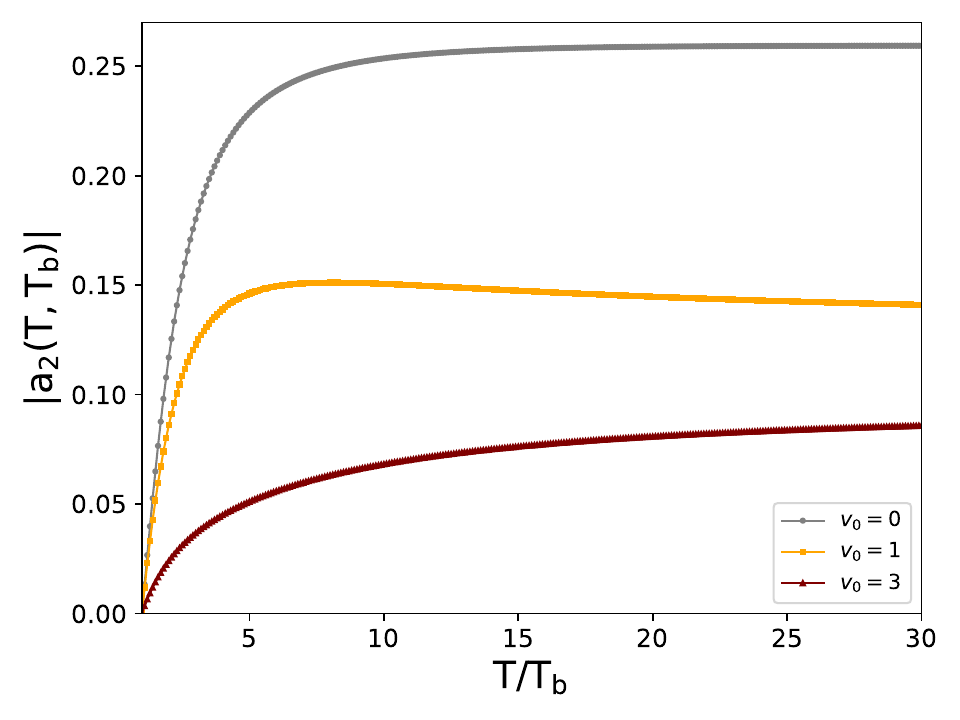}
\caption{\label{a2 vs T UL 6}Presence of activity induces Mpemba effect although further increase in activity suppresses the effect. Variation of $|a_2(T,T_b)|$ with $T$ for the chosen set of parameters: $U_{\ell}$=6, $U_r$=10, $\gamma=1$, $\tau_p=1$ and for the choices of the persistence velocities: $v_0=0$, $v_0=1$ and $v_0=3$. The final quenched temperature corresponds to $T_b=1$. Compared to the passive case ($v_0=0$) where there is no Mpemba effect, it emerges as $v_0$ is increased but it vanishes with further increase in the activity.}
\end{figure}

In order to verify if the role of activity in the Mpemba effect as appeared in the above example is universal, we next consider the opposite case where the Mpemba effect is originally absent for the passive  particle ($v_0$=0). In our example, it corresponds to making a choice of the external potential $U_{\ell}=6.0$, $U_r=10.0$ and $\alpha=1.0$ with $T_b=1.0$. The absence of the Mpemba effect is evident from the monotonic rise in $|a_2(T,T_b)|$ with $T$ for $v_0=0$ as shown in \fref{a2 vs T UL 6}. However, in contrast to the previous case where activity above a certain threshold  suppressing the Mpemba effect, on the introduction of the active degree of freedom of the particle characterized by non-zero $v_0$, the system shows the Mpemba effect as evident from the non-monotonic rise of $|a_2(T,T_b)|$ with $T$ for $v_0=1$ [see \fref{a2 vs T UL 6}]. Thus, in this case, we observe an opposite behavior of the activity enhancing the Mpemba effect as compared to the previous case. However, on further increasing the activity of the system to $v_0=3$, the Mpemba effect vanishes and the relaxation behavior turns out to be the same as in the passive system.

Concluding from the behavior of activity in the two examples considered, it can be inferred that large persistence velocity of the Brownian particle always suppresses the Mpemba effect. In effect, increasing the persistence velocity of the particle overcome any source of slowness in the dynamics of the system arising from the external potential leading to no Mpemba effect.

\section{\label{phase diagram}Phase diagram}
We have seen that having an intermediate value of activity leads to the Mpemba effect where it is absent in the passive case. On the contrary, for the case where the Mpemba effect is already present in the passive model, introducing activity above a certain threshold suppresses the Mpemba effect. However, in the presence of large activity, the Mpemba effect is always absent irrespective of whether it is present or absent in the passive model. In order to have a broader understanding about the role of activity in the Mpemba effect, it can be more informative to seek how the presence of activity affects in general the phase diagram of system parameters that leads to the Mpemba effect for the passive model.

To that end, we study the different configurations of the potential well to differentiate which configurations allow and do not allow the Mpemba like relaxations, as the activity is varied starting from the scenario where $v_0=0$ that corresponds to the phase diagram of the passive model. This allows us to understand the pattern in which the phase diagram of the passive model changes with the introduction of activity.  

\begin{figure}
\centering
\includegraphics[width= \columnwidth]{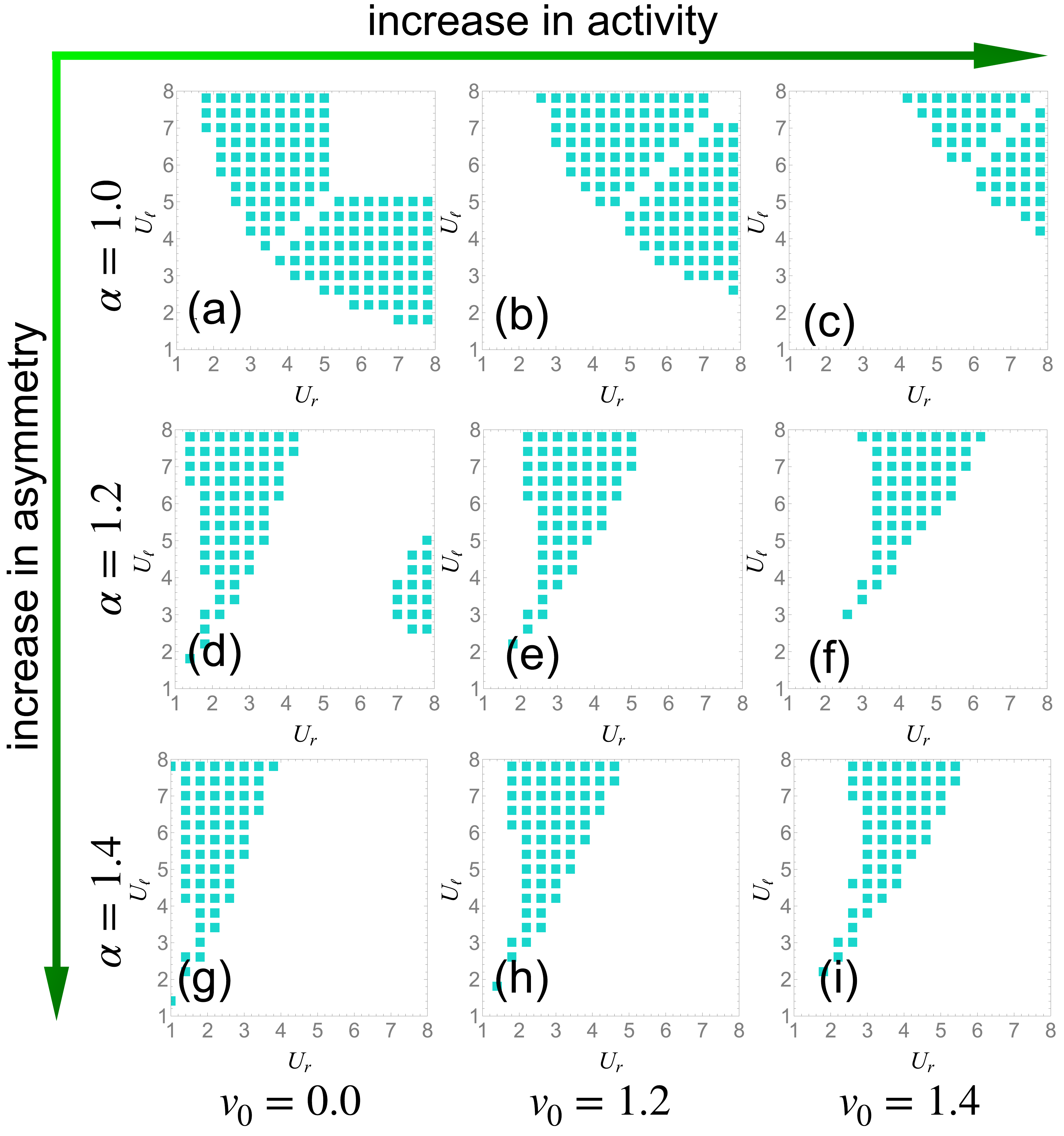} 
\caption{\label{phase diag}Change in the phase space region of initial conditions that show the Mpemba effect (shown as coloured region) as a function of the change in activity characterized by persistence velocity, $v_0$ and change in asymmetry parameter, $\alpha$. The final quenched temperature corresponds to $T_b=1$.}
\end{figure}

Note that the configuration of the potential well depends on the three parameters: $U_{\ell}$, $U_r$ and $\alpha$. From the study of the Mpemba effect for passive colloids in the presence of single well potential, it is known that the asymmetry in the potential $U(x)$ is the key and it can be introduced through $U_{\ell}\neq U_r$ and/ or $\alpha\neq 1$. In a similar vein, we explore the effect of all three parameters and also study the change in behavior of the phase diagram as the activity is varied starting from the passive model.  For that purpose, we determine the phase diagram in the $U_\ell$--$U_r$ plane for different $\alpha$ and $v_0$ as shown in Fig.~\ref{phase diag}.

For the passive model ($v_0=0$), the symmetric case $U_{\ell}=U_r$ does not show the Mpemba effect even if $\alpha\neq 1$ [see Ref.~\cite{PhysRevE.108.024131} and Figs.~\ref{phase diag}(d), (g)]. The same is true even in the presence of activity ($v_0\neq 0$) and thus a minimal asymmetry in $U_{\ell}$ and $U_r$ is the key to observe the Mpemba effect. Next, having known that asymmetry in $U_{\ell}$ and $U_r$ is important, we now probe the behavior of the phase space region  as the parameters $\alpha$ and $v_0$ are varied. 

We know the effect of the asymmetry parameter $\alpha$ for the passive model, that it decreases the phase space of the Mpemba effect as $\alpha$ is increased. Now in the same study we seek to understand how the phase space behaves as the activity is varied. We find that with the simultaneous increase in activity along with $\alpha$ affects the phase space in two ways: first, unlike the passive case, the increase in $\alpha$ can either decrease or increase the phase space region depending on the value of the activity. As shown in Fig.~\ref{phase diag}, the phase space region decreases with the increase in $\alpha$ for $v_0=1.2$ while it increases for $v_0=1.4$. Thus, in the presence of activity, the role of asymmetry in the increase or decrease of the phase space seems ambiguous. However, for a given $\alpha$, the increase in activity effectively shifts the phase space of the passive model. As a result, it leads to the emergence (and depletion) of the phase space region that shows the Mpemba effect in the presence of activity where it was initially absent (or present) for the passive case. Thus, it leads to the activity induced and suppressed Mpemba effect.


 \section{\label{conclusion}Conclusion}
In summary, we studied the Mpemba effect for an active Brownian particle in a single well potential that is piecewise linear. 
While the presence of the Mpemba effect in the setup of a single well potential with a passive Brownian particle had been confirmed earlier where the relaxation starts from an equilibrium Boltzmann distribution, the presence of activity presents a non-equilibrium steady state and the consequences of the presence of the activity in the relaxation process was the main motive of the current study.
To that end, we explored how the phase space pertaining to different configurations of the external potential that leads to the Mpemba effect is affected with the change in activity parameters.


We first show that the presence of activity can show the Mpemba effect for a given parameter space where it was absent for the passive Brownian particle and vice-versa leading to the activity induced and suppressed Mpemba effect compared to the system of passive colloid. The change in the pattern of the phase space regions that show the Mpemba effect, with the increase in activity  sheds light  on the activity induced and suppressed Mpemba effect.

While the phase space that shows the Mpemba effect changes in the presence of the activity, the necessity of a minimal asymmetry in $U_{\ell}$ and $U_r$ of the external potential for the Mpemba effect remains unchanged as in the passive case. But unlike the phase space of the Mpemba effect for the passive case that decreases with the increase in additional asymmetry in terms of parameter $\alpha$ of the potential well, the region of phase space showing the Mpemba effect can increase or decrease with $\alpha$ depending on the value of the activity. Moreover, for a given value of the asymmetry parameter $\alpha$, the increase in activity leads to an effective translational shift in the region of phase space that shows the Mpemba effect when compared to the passive case. As a result, addition and deletion in the existing phase space of the passive setup takes place and hence it clarifies the fact of why the activity induced Mpemba effect takes place for a set of system parameters where it is absent in the passive case and vice-versa.  

It would be interesting to explore such anomalous relaxations in experiments as setups of trapped active colloids are ubiquitous. Another direction of future research using the current setup would be to explore the time delayed cooling protocols~\cite{santos2024mpemba} to investigate the emergence of the Mpemba and Kovacs effect.

\section*{Data Availability Statement}
The data that support the findings of this study are available upon reasonable request from the authors.

%

\end{document}